\documentclass[aps,preprint]{revtex4}%
\usepackage{amsfonts}
\usepackage{amsmath}
\usepackage{amssymb}
\usepackage{graphicx}%
\begin{document}
\preprint{CTP-SCU/2020001}
\title{Chaotic Motion around a Black Hole under Minimal Length Effects}
\author{Xiaobo Guo$^{a}$}
\email{guoxiaobo@czu.edu.cn}
\author{Kangkai Liang$^{c,d}$}
\email{lkk@berkeley.edu}
\author{Benrong Mu$^{b}$}
\email{benrongmu@cdutcm.edu.cn}
\author{Peng Wang$^{c}$}
\email{pengw@scu.edu.cn}
\author{Mingtao Yang$^{c}$}
\email{2017141221040@stu.scu.edu.cn}
\affiliation{$^{a}$ Mechanical and Electrical Engineering School, Chizhou University,
Chizhou, 247000, PR China}
\affiliation{$^{b}$Physics Teaching and Research section, College of Medical Technology,
Chengdu University of Traditional Chinese Medicine, Chengdu, 611137, PR China}
\affiliation{$^{c}$Center for Theoretical Physics, College of Physics, Sichuan University,
Chengdu, 610064, PR China}
\affiliation{$^{d}$Department of Physics, University of California at Berkeley, Berkeley,
CA, 94720, USA}

\begin{abstract}
We use the Melnikov method to identify chaotic behavior in geodesic motion
perturbed by the minimal length effects around a Schwarzschild black hole.
Unlike the integrable unperturbed geodesic motion, our results show that the
perturbed homoclinic orbit, which is a geodesic joining the unstable circular
orbit to itself, becomes chaotic in the sense that Smale horseshoes chaotic
structure is present in phase space.

\end{abstract}
\keywords{}\maketitle
\tableofcontents

\bigskip

%\affiliation{Center for Theoretical Physics, College of Physical Science and Technology,
%Sichuan University, Chengdu, 610064, PR China}

%\affiliation{Center for Theoretical Physics, College of Physical Science and Technology,
%Sichuan University, Chengdu, 610064, PR China}

\section{Introduction}

Chaos is now one of the most important ideas to understand various nonlinear
phenomena in general relativity. Chaos in geodesic motion can lead to
astrophysical applications and provide some important insight into AdS/CFT
correspondence. However, the geodesic motion of a point particle in the
generic Kerr-Newman black hole spacetime is well known to be integrable
\cite{IN-Carter:1968rr}, which leads to the absence of chaos. So complicated
geometries of spacetime or extra forces imposed upon the particle are
introduced to study the chaotic geodesic motion of a test particle. Examples
of chaotic behavior of geodesic motion of particles in various backgrounds
were considered in
\cite{IN-Sota:1995ms,IN-Hanan:2006uf,IN-Gair:2007kr,IN-Witzany:2015yqa,IN-Wang:2016wcj,IN-Chen:2016tmr,IN-Liu:2018bmn}%
. On the other hand, the geodesic motion of a ring string instead of a point
particle has been shown to exhibit chaotic behavior in a Schwarzschild black
hole \cite{IN-Frolov:1999pj}. Later, the chaotic dynamics of ring strings was
studied in other black hole backgrounds
\cite{IN-Zayas:2010fs,IN-Ma:2014aha,IN-Ma:2019ewq}.

Among the various indicators for detecting chaos, the Melnikov method is an
analytical approach applicable to near integrable perturbed systems and has as
its main advantages the fact that knowledge of the unperturbed integrable
dynamics is only required \cite{IN-Mel}. The Melnikov method has been used to
discuss the chaotic behavior of geodesic motion in black holes perturbed by
gravitational waves \cite{IN-Bombelli:1991eg,IN-Letelier:1996he},
electromagnetic fields \cite{IN-Santoprete:2001wz} and a thin disc
\cite{IN-Polcar:2019kwu}. Recently, chaos\ due to temporal and spatially
periodic perturbations in charged AdS black holes has also been investigated
via the Melnikov method
\cite{IN-Chabab:2018lzf,IN-Mahish:2019tgv,IN-Chen:2019bwt,IN-Dai:2020wny}.

The existence of a minimal measurable length has been observed in various
quantum theories of gravity such as string theory
\cite{IN-Veneziano:1986zf,IN-Gross:1987ar,IN-Amati:1988tn,IN-Garay:1994en}.
The generalized uncertainty principle (GUP) was proposed to incorporate the
minimal length into quantum mechanics
\cite{IN-Maggiore:1993kv,IN-Kempf:1994su}. The GUP can lead to the minimal
length deformed fundamental commutation relation. For a $1$D quantum system,
the deformed commutator between position and momentum can take the following
form
\begin{equation}
\lbrack X,P]=i\hbar(1+\beta P^{2}), \label{eq:1dGUP}%
\end{equation}
where $\beta$ is some deformation parameter, and the minimal length is $\Delta
X_{\min}=\hbar\sqrt{\beta}$. In the context of the minimal length deformed
quantum mechanics, various quantum systems have been investigated intensively,
e.g. the harmonic oscillator \cite{IN-Chang:2001kn}, Coulomb potential
\cite{IN-Akhoury:2003kc,IN-Brau:1999uv}, gravitational well
\cite{IN-Brau:2006ca,IN-Pedram:2011xj}, quantum optics
\cite{IN-Pikovski:2011zk,IN-Bosso:2018ckz} and compact stars
\cite{IN-Wang:2010ct,IN-Ong:2018zqn}. In the classical limit $\hbar
\rightarrow0$, the effects of the minimal length can be studied in the
classical context. For example, the minimal length effects have been analyzed
for the observational tests of general relativity
\cite{IN-Benczik:2002tt,IN-Ahmadi:2014cga,IN-Silagadze:2009vu,IN-Scardigli:2014qka,IN-Ali:2015zua,IN-Guo:2015ldd,IN-Khodadi:2017eim,IN-Scardigli:2018qce}%
, classical harmonic oscillator \cite{IN-Tao:2012fp,IN-Quintela:2015bua},
equivalence principle \cite{IN-Tkachuk:2013qa}, Newtonian potential
\cite{IN-Scardigli:2016pjs}, the Schroinger-Newton equation
\cite{IN-Zhao:2017xjj}, the weak cosmic censorship conjecture
\cite{IN-Mu:2019bim} and motion of particles near a black hole horizon
\cite{IN-Lu:2018mpr,IN-Hassanabadi:2019iff}. Moreover, the minimal length
corrections to the Hawking temperature were also obtained using the
Hamilton-Jacobi method in
\cite{IN-Chen:2013tha,IN-Chen:2013ssa,IN-Chen:2014xgj,IN-Maghsoodi:2019fca}.

In \cite{IN-Lu:2018mpr}, we considered the minimal length effects on motion of
a massive particle near the black hole horizon under some external potential,
which was introduced to put the particle at the unstable equilibrium outside
the horizon. It was found that the minimal length effects could make the
classical trajectory in black holes more chaotic, which motivates us to study
the minimal length effects on geodesic motion in black holes. In this paper,
we use the Melnikov method to investigate the homoclinic orbit perturbed by
the minimal length effects in a Schwarzschild black hole and find that the
perturbed homoclinic orbit breaks up into a chaotic layer. For simplicity, we
set $\hbar=c=G=1$ in this paper.

\section{Melnikov Method}

\label{Sec:MM}

The Melnikov method provides a tool to determine the existence of chaos in
some dynamical system under nonautonomous periodic perturbations. The
existence of simple zeros of the Melnikov function leads to the Smale
horseshoes structure in phase space, which implies that the dynamical system
is chaotic. In this section, we briefly review the classical Melnikov method
and the generalization of the Melnikov method in a system with two coordinate
variables, one of which is periodic. Note that \cite{IN-Bombelli:1991eg}
provides a concise introduction to the Melnikov method.

The classical Melnikov method is applied to a dynamical system with one degree
of freedom, whose Hamiltonian is given by%
\begin{equation}
\mathcal{H}\left(  p,q,t\right)  =\mathcal{H}_{0}\left(  p,q\right)
+\epsilon\mathcal{H}_{1}\left(  p,q,t\right)  .
\end{equation}
Here $\mathcal{H}_{0}\left(  p,q\right)  $ describes an unperturbed integrable
system, $\mathcal{H}_{1}\left(  p,q,t\right)  $ is a nonautonomous periodic
perturbation of $t$ with some period $T$, and the small parameter $\epsilon$
controls the perturbation. Moreover, we assume $\mathcal{H}_{0}\left(
p,q\right)  $ contains a hyperbolic fixed point $\left(  q_{0},p_{0}\right)  $
and a homoclinic orbit $\left(  q_{0}\left(  t\right)  ,p_{0}\left(  t\right)
\right)  $ corresponding to this fixed point. The homoclinic orbit $\left(
q_{0}\left(  t\right)  ,p_{0}\left(  t\right)  \right)  $ joins $\left(
q_{0},p_{0}\right)  $ to itself:
\begin{equation}
\left(  q_{0}\left(  t\right)  ,p_{0}\left(  t\right)  \right)  \rightarrow
\left(  q_{0},p_{0}\right)  \text{ as }t\rightarrow\pm\infty\text{.}%
\end{equation}
Roughly speaking, the stable/unstable manifold of a fixed point consists of
points that approach the fixed point in the limit of $t\rightarrow
+\infty/t\rightarrow-\infty$. In the unperturbed system, the stable manifold
of $\left(  q_{0},p_{0}\right)  $ coincides with the unstable manifold along
the homoclinic orbit $\left(  q_{0}\left(  t\right)  ,p_{0}\left(  t\right)
\right)  $. When the perturbation is switched on, the fixed point $\left(
q_{0},p_{0}\right)  $ becomes a single periodic orbit $\left(  q^{\epsilon
}\left(  t\right)  ,p^{\epsilon}\left(  t\right)  \right)  $ with period $T$
around $\left(  q_{0},p_{0}\right)  $. Choosing an arbitrary initial time
$t_{0}$, we can define the Poincare map $\phi_{t_{0}}$, which maps a point in
the phase space to its image after $T$ along the flow of the perturbed
Hamiltonian. Under the Poincare map $\phi_{t_{0}}$, $\left(  q^{\epsilon
}\left(  t_{0}\right)  ,p^{\epsilon}\left(  t_{0}\right)  \right)  $ is a
fixed point, and the stable and unstable manifolds of this fixed point usually
do not coincide. The distance between these manifolds measured along a
direction that is perpendicular to the unperturbed homoclinic orbit $\left(
q_{0}\left(  t\right)  ,p_{0}\left(  t\right)  \right)  $ is proportional to
the Melnikov function \cite{MM-Wiggins},%
\begin{equation}
\mathcal{M}\left(  t_{0}\right)  =\int_{-\infty}^{+\infty}\left\{
\mathcal{H}_{0},\mathcal{H}_{1}\right\}  \left(  q_{0}\left(  t\right)
,p_{0}\left(  t\right)  ,t_{0}+t\right)  dt,
\end{equation}
where $\left\{  \mathcal{\ },\mathcal{\ }\right\}  $ is the Poisson bracket.
It has been shown \cite{MM-Wiggins} that when $\mathcal{M}\left(
t_{0}\right)  $ has a simple zero, i.e., $\mathcal{M}\left(  t_{0}\right)  =0$
and $d\mathcal{M}\left(  t_{0}\right)  /dt_{0}\neq0$, the stable and unstable
manifolds intersect transversally, which leads to a homoclinic tangle and
consequently Smale horseshoes. The presence of Smale horseshoes means the
orbit turns into a chaotic layer.

In \cite{IN-Polcar:2019kwu}, the Melnikov method was extended to a
two-degrees-of-freedom system with the Hamiltonian%
\begin{equation}
\mathcal{H}\left(  p,q,\psi,J\right)  =\mathcal{H}_{0}\left(  p,q,J\right)
+\epsilon\mathcal{H}_{1}\left(  p,q,\psi,J\right)  ,
\end{equation}
where the coordinate $\psi$ is periodic, and $J$ is its conjugate momentum.
The Hamiltonian of the system does not depend explicitly on time, and $\psi$
can play the role of time. For the unperturbed system, using the equation of
motion $\dot{\psi}=\partial\mathcal{H}_{0}\left(  p,q,J\right)  /\partial J$,
we can express the homoclinic orbit in terms of $\psi$, i.e., $\left(
q_{0}\left(  \psi\right)  ,p_{0}\left(  \psi\right)  \right)  $. Here, the dot
denotes derivative with respect to $t$. Holmes \& Marsden showed \cite{MM-HM}
that the Melnikov function in this two-degrees-of-freedom system is given by%
\begin{equation}
\mathcal{M}\left(  \psi_{0}\right)  =\int_{-\infty}^{\infty}\frac{1}{\dot
{\psi}\left(  q_{0}\left(  \psi\right)  ,p_{0}\left(  \psi\right)  ,J\right)
}\left\{  \mathcal{H}_{0},\frac{\mathcal{H}_{1}}{\dot{\psi}}\right\}  \left(
q_{0}\left(  \psi\right)  ,p_{0}\left(  \psi\right)  ,\psi_{0}+\psi,J\right)
d\psi, \label{eq:Mfunction}%
\end{equation}
where the Poisson bracket is only computed in terms of $q$ and $p$. The
Melnikov function $\mathcal{M}\left(  \psi_{0}\right)  $ is periodic and has
the same period as $\psi$. When $\mathcal{M}\left(  \psi_{0}\right)  $ has a
simple zero, the perturbation $\mathcal{H}_{1}$ makes the system chaotic.

\section{Chaos Under Minimal Length Effects}

\label{Sec:CUMLE}

In this section, we use the Melnikov method to investigate the chaotic
dynamics of particles around a Schwarzschild black hole under the minimal
length effects. The Schwarzschild metric is
\begin{equation}
ds^{2}=g_{\mu\nu}dx^{\mu}dx^{\nu}=-f\left(  r\right)  dt^{2}+\frac{dr^{2}%
}{f\left(  r\right)  }+r^{2}\left(  d\theta^{2}+\sin^{2}\theta d\phi
^{2}\right)  ,
\end{equation}
where $f\left(  r\right)  =1-2M/r$, and $M$ is the black hole mass. There are
various ways to study the geodesic motion of a particle around a black hole.
Specifically, the geodesics can be obtained using the Hamilton-Jacobi method.
In \cite{IN-Guo:2015ldd,CUMLE-Mu:2015qta}, the minimal length deformed
Hamilton-Jacobi equation in a spherically symmetric black hole background was
derived by taking the WKB limit of the deformed Klein-Gordon, Dirac and
Maxwell's equations. For the deformed fundamental commutation relation
$\left(  \ref{eq:1dGUP}\right)  $, the deformed Hamilton-Jacobi equation for a
massive relativistic point particle is
\begin{equation}
E_{\mathcal{S}}^{\left(  0\right)  }+2\beta E_{\mathcal{S}}^{\left(  1\right)
}=0,
\end{equation}
where
\begin{align}
E_{\mathcal{S}}^{\left(  0\right)  }  &  \equiv-\frac{\left(  \partial
_{t}S\right)  ^{2}}{f\left(  r\right)  }+f\left(  r\right)  \left(
\partial_{r}S\right)  ^{2}+\frac{\left(  \partial_{\theta}S\right)  ^{2}%
}{r^{2}}+\frac{\left(  \partial_{\phi}S\right)  ^{2}}{r^{2}\sin^{2}\theta
}+m^{2},\nonumber\\
E_{\mathcal{S}}^{\left(  1\right)  }  &  \equiv-\frac{\left(  \partial
_{t}S\right)  ^{4}}{f^{2}\left(  r\right)  }+f^{2}\left(  r\right)  \left(
\partial_{r}S\right)  ^{4}+\frac{\left(  \partial_{\theta}S\right)  ^{4}%
}{r^{4}}+\frac{\left(  \partial_{\phi}S\right)  ^{4}}{r^{4}\sin^{4}\theta},
\end{align}
and $S$ is the classical action. There are no explicit $t$- and $\phi
$-dependence in the Hamilton-Jacobi equation, so $S$ is separable,
\begin{equation}
S=S_{r}\left(  r\right)  +S_{\theta}\left(  \theta\right)  -mEt+mL_{\phi}\phi,
\end{equation}
where $E$ and $L_{\phi}$ have the meaning of the energy per unit mass and
$z$-component of the orbital angular momentum per unit mass, respectively.
Since $p_{\mu}=\partial_{\mu}S$, we can rewrite $E_{\mathcal{S}}^{\left(
0\right)  }$ and $E_{\mathcal{S}}^{\left(  1\right)  }$ as%
\begin{align}
E_{\mathcal{S}}^{\left(  0\right)  }\left(  r,p_{r},\theta,p_{\theta}\right)
&  \equiv-\frac{m^{2}E^{2}}{f\left(  r\right)  }+f\left(  r\right)  p_{r}%
^{2}+\frac{p_{\theta}^{2}}{r^{2}}+\frac{m^{2}L_{\phi}^{2}}{r^{2}\sin^{2}%
\theta}+m^{2},\nonumber\\
E_{\mathcal{S}}^{\left(  1\right)  }\left(  r,p_{r},\theta,p_{\theta}\right)
&  \equiv-\frac{m^{4}E^{4}}{f^{2}\left(  r\right)  }+f^{2}\left(  r\right)
p_{r}^{4}+\frac{p_{\theta}^{4}}{r^{4}}+\frac{m^{4}L_{\phi}^{4}}{r^{4}\sin
^{4}\theta},
\end{align}
where $p_{\mu}$ are the conjugate momentums.

The unperturbed Hamilton-Jacobi equation $E_{\mathcal{S}}^{\left(  0\right)
}=0$ describes the geodesic motion of a particle around a Schwarzschild black
hole in the usual case without the minimal length effects. To find the
unperturbed Hamiltonian, we start with the Lagrangian for a massive
relativistic point particle%
\begin{equation}
\mathcal{L=}\frac{g_{\mu\nu}}{2e}\frac{dx^{\mu}}{d\tau}\frac{dx^{\mu}}{d\tau
}-\frac{em^{2}}{2},
\end{equation}
where $\tau$ is the world-line parameter, and $e$ is an einbein field. The
corresponding Hamiltonian is%
\begin{equation}
\mathcal{H}_{0}=\frac{dx^{\mu}}{d\tau}\frac{\partial\mathcal{L}}%
{\partial\left(  dx^{\mu}/d\tau\right)  }-\mathcal{L=}\frac{e}{2}\left(
g^{\mu\nu}p_{\mu}p_{\nu}+m^{2}\right)  ,
\end{equation}
which is just $E_{\mathcal{S}}^{\left(  0\right)  }$ if we choose $e=2$. Using
the equations of motion, one can obtain the Hamiltonian constraint
$\mathcal{H}_{0}=0$, which is a consequence of the gauge symmetry associated
with the reparameterization symmetry of $\tau$. It is worth noting that the
Hamiltonian constraint $\mathcal{H}_{0}=0$ is precisely the Hamilton-Jacobi
equation $E_{\mathcal{S}}^{\left(  0\right)  }=0$.

The unperturbed Hamilton-Jacobi equation $E_{\mathcal{S}}^{\left(  0\right)
}=0$ further leads to
\begin{align}
\frac{p_{\theta}}{m}  &  =\sqrt{L^{2}-\frac{L_{\phi}^{2}}{\sin^{2}\theta}%
},\nonumber\\
\frac{\left(  p^{r}\right)  ^{2}}{m^{2}}  &  =\left(  \frac{dr}{d\tau}\right)
^{2}=E^{2}-V_{\text{eff}}\left(  u\right)  , \label{eq:pthetapr}%
\end{align}
where $L$ is the angular momentum per unit mass, $V_{\text{eff}}\left(
u\right)  \equiv\left(  1-2u\right)  \left(  1+u^{2}L^{2}/M^{2}\right)  $ is
the effective potential, and $u\equiv M/r$. The radius $u_{f}$ and the energy
$E_{f}$ of the unstable circular orbit are determined by $dV_{\text{eff}%
}/du=0$ and $d^{2}V_{\text{eff}}/du^{2}<0$, which gives%
\begin{equation}
u_{f}=\frac{1+\sqrt{1-12M^{2}/L^{2}}}{6}\text{ and }E_{f}=\sqrt{\left(
1-2Mu_{f}\right)  \left(  1+L^{2}u_{f}^{2}\right)  }.
\end{equation}
The hyperbolic fixed point of $\mathcal{H}_{0}$ in the $u$-$p_{r}$ phase space
is $\left(  u_{f},0\right)  $. Since $p_{\theta}$ is not an integral of motion
for $\mathcal{H}_{0}$, a new pair of action-angle like variables $J$ and
$\psi$ were introduced to make the Melnikov method applicable
\cite{IN-Polcar:2019kwu}. In fact, $J$ and $\psi$ are given by%
\begin{align}
J  &  \equiv\frac{1}{\pi}\int_{\arcsin(L_{\phi}/L)}^{\pi-\arcsin(L_{\phi}%
/L)}p_{\theta}d\theta=m\left(  L-L_{\phi}\right)  ,\nonumber\\
\psi &  =\frac{\partial S_{\theta}}{\partial J}=\frac{1}{m}\int\frac{\partial
p_{\theta}}{\partial L}d\theta=-\arctan\left(  \frac{\cos\theta}{\sqrt
{\sin^{2}\theta-L_{\phi}^{2}/L^{2}}}\right)  , \label{eq:Jandphi}%
\end{align}
which shows that $\psi$ is periodic, and the period is $\pi$.

The homoclinic orbit $u_{0}\left(  \psi\right)  $ connecting $u_{f}$ to itself
has the same energy $E_{f}$ as the unstable circular orbit and is determined
by%
\begin{equation}
\frac{du_{0}\left(  \psi\right)  }{d\psi}=\frac{M\sqrt{E_{f}^{2}%
-V_{\text{eff}}\left(  u\right)  }}{L},
\end{equation}
where we use $dr/d\tau=-Ldu/\left(  Md\psi\right)  $ and eqn. $\left(
\ref{eq:pthetapr}\right)  $. Integrating the above equation, one has%
\begin{equation}
u_{0}\left(  \psi\right)  =u_{m}+\left(  u_{f}-u_{m}\right)  \tanh^{2}\left(
\sqrt{\frac{u_{f}-u_{m}}{2}}\psi\right)  , \label{eq:homorbit}%
\end{equation}
where $u_{m}=1/2-2u_{f}$ and $u_{0}\left(  \psi\rightarrow\pm\infty\right)
=u_{f}$. Note that the existence of the homoclinic orbit $u_{0}\left(
\psi\right)  $ requires that $u_{f}>u_{m}>0$, which gives $2\sqrt{3}M<L<4M$.

In the unperturbed case, we show that the Hamilton-Jacobi equation can be
interpreted as the Hamiltonian constraint, which means $\mathcal{H}%
_{0}=E_{\mathcal{S}}^{\left(  0\right)  }$. Similarly, in the perturbed case,
we can also treat the Hamilton-Jacobi equation $E_{\mathcal{S}}^{\left(
0\right)  }+2\beta E_{\mathcal{S}}^{\left(  1\right)  }=0$ as the Hamiltonian
constraint, which leads to the perturbed Hamiltonian $\mathcal{H}$,%
\begin{equation}
\mathcal{H}=E_{\mathcal{S}}^{\left(  0\right)  }+2\beta E_{\mathcal{S}%
}^{\left(  1\right)  }\text{.}%
\end{equation}
Taking $\epsilon=2\beta$, the perturbation $\mathcal{H}_{1}$ is then given by%
\begin{equation}
\mathcal{H}_{1}=E_{\mathcal{S}}^{\left(  1\right)  }.
\end{equation}
Using eqn. $\left(  \ref{eq:Jandphi}\right)  $, we can express $\mathcal{H}%
_{0}$ and $\mathcal{H}_{1}$ as functions of $r$, $p_{r}$, $\psi$ and $J$,
\begin{align}
\mathcal{H}_{0}\left(  r,p_{r},J\right)   &  =-\frac{m^{2}E^{2}}{f\left(
r\right)  }+f\left(  r\right)  p_{r}^{2}+\frac{m^{2}L^{2}}{r^{2}}%
+m^{2},\nonumber\\
\mathcal{H}_{1}\left(  r,p_{r},\psi,J\right)   &  =-\frac{m^{4}E^{4}}%
{f^{2}\left(  r\right)  }+f^{2}\left(  r\right)  p_{r}^{4}+\frac{m^{4}}{r^{4}%
}\left(  L^{2}-\frac{L_{\phi}^{2}}{A\left(  \psi\right)  }\right)  ^{2}%
+\frac{m^{4}L_{\phi}^{4}}{r^{4}A^{2}\left(  \psi\right)  }, \label{eq:H0H1}%
\end{align}
where $A\left(  \psi\right)  \equiv1+\left(  L_{\phi}^{2}/L^{2}-1\right)
\sin^{2}\psi$ and $L=J/m+L_{\phi}$.

Substituting into eqn. $\left(  \ref{eq:Mfunction}\right)  $ the homoclinic
orbit $\left(  \ref{eq:homorbit}\right)  $ and the corresponding conjugate
momentum as%
\begin{equation}
r_{0}\left(  \psi\right)  =\frac{M}{u_{0}\left(  \psi\right)  }\text{ and
}p_{0}^{r}\left(  \psi\right)  =-\frac{mL}{M}\frac{du_{0}\left(  \psi\right)
}{d\psi},
\end{equation}
the Melnikov function becomes%
\begin{equation}
\mathcal{M}\left(  \psi_{0}\right)  =\int_{-\infty}^{\infty}\frac{M^{2}%
}{2\left(  J/m+L_{\phi}\right)  u_{0}^{2}\left(  \psi\right)  }\left\{
\mathcal{H}_{0},\frac{\mathcal{H}_{1}}{\dot{\psi}}\right\}  \left(  \frac
{M}{u_{0}\left(  \psi\right)  },-\frac{mL}{M}\frac{du_{0}\left(  \psi\right)
}{d\psi},\psi_{0}+\psi,J\right)  d\psi, \label{eq:M0}%
\end{equation}
where we use
\begin{equation}
\dot{\psi}=\frac{\partial\mathcal{H}_{0}}{\partial J}=\frac{2\left(
J/m+L_{\phi}\right)  u_{0}^{2}\left(  \psi\right)  }{M^{2}}.
\end{equation}
We find that $\mathcal{M}\left(  \psi_{0}\right)  $ can be rewritten as
\begin{equation}
\mathcal{M}\left(  \psi_{0}\right)  =\beta m^{5}Mh\left(  \psi_{0},\frac{L}%
{M},\frac{L_{\phi}}{M}\right)  , \label{eq:M}%
\end{equation}
where $h$ is some function of the dimensionless variables $\psi_{0},L/M$ and
$L_{\phi}/M$.

\begin{figure}[tb]
\begin{center}
\includegraphics[width=0.55\textwidth]{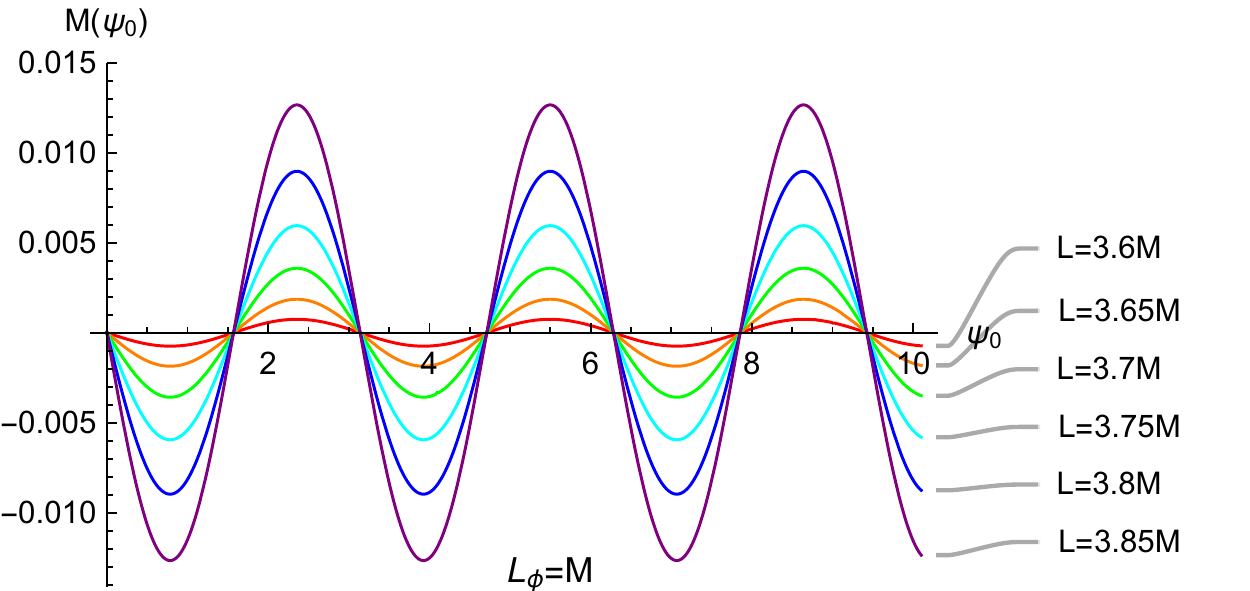}
\includegraphics[width=0.42\textwidth]{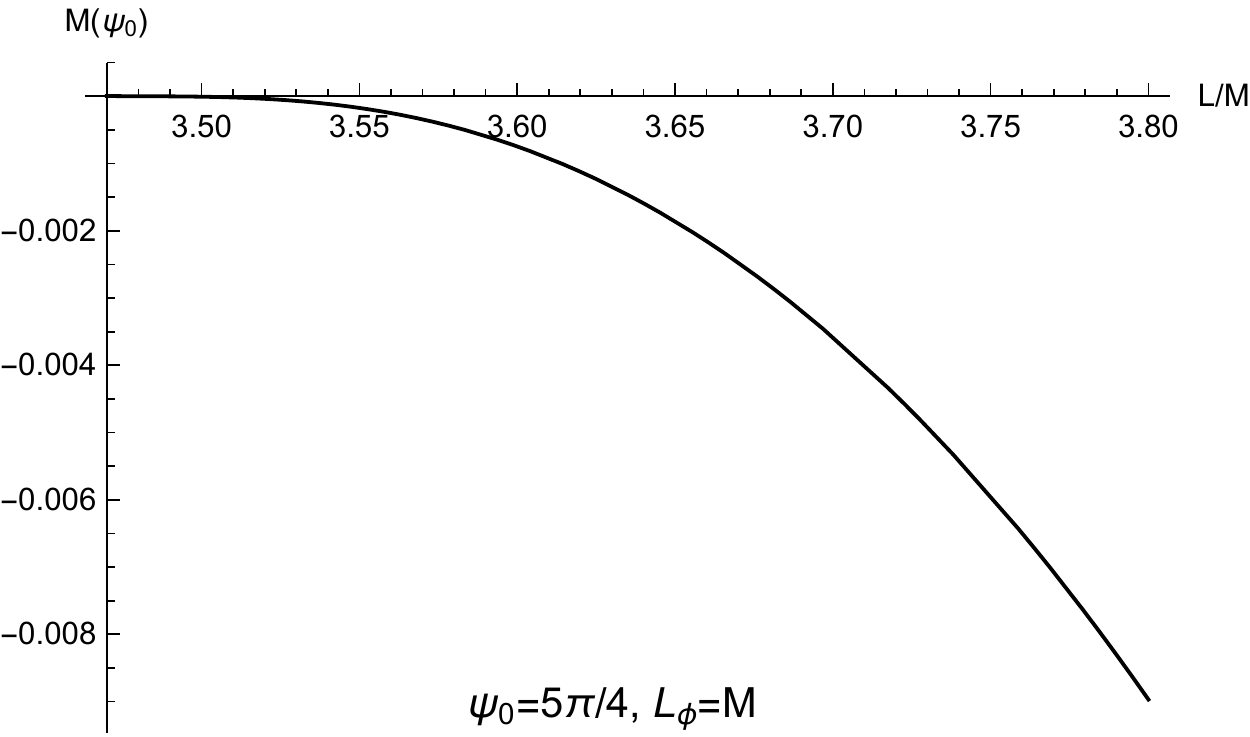}
\end{center}
\caption{{\footnotesize Dependence of the Melnikov function $\mathcal{M}%
\left(  \psi_{0}\right)  $ on the angular momentum $L$. We take $m=1$ and
$L_{\phi}=M$. \textbf{Left Panel}: $\mathcal{M}\left(  \psi_{0}\right)  $ as a
function of $\psi_{0}$ for various values of $L$. It shows that $\mathcal{M}%
\left(  \psi_{0}\right)  $ is a periodic function with the period of $\pi$,
and $\mathcal{M}\left(  \psi_{0}\right)  $ has simple zeros at the points
$\psi_{0}=n\pi/2$ with $n\in Z$. \textbf{Right Panel}: $\mathcal{M}\left(
\psi_{0}=5\pi/4\right)  $ as a function of $L/M$ on the interval of $2\sqrt
{3}<L/M<4$. The amplitude of $\mathcal{M}\left(  \psi_{0}\right)  $ increases
with increasing $L$.}}%
\label{fig:L}%
\end{figure}

\begin{figure}[tb]
\begin{center}
\includegraphics[width=0.55\textwidth]{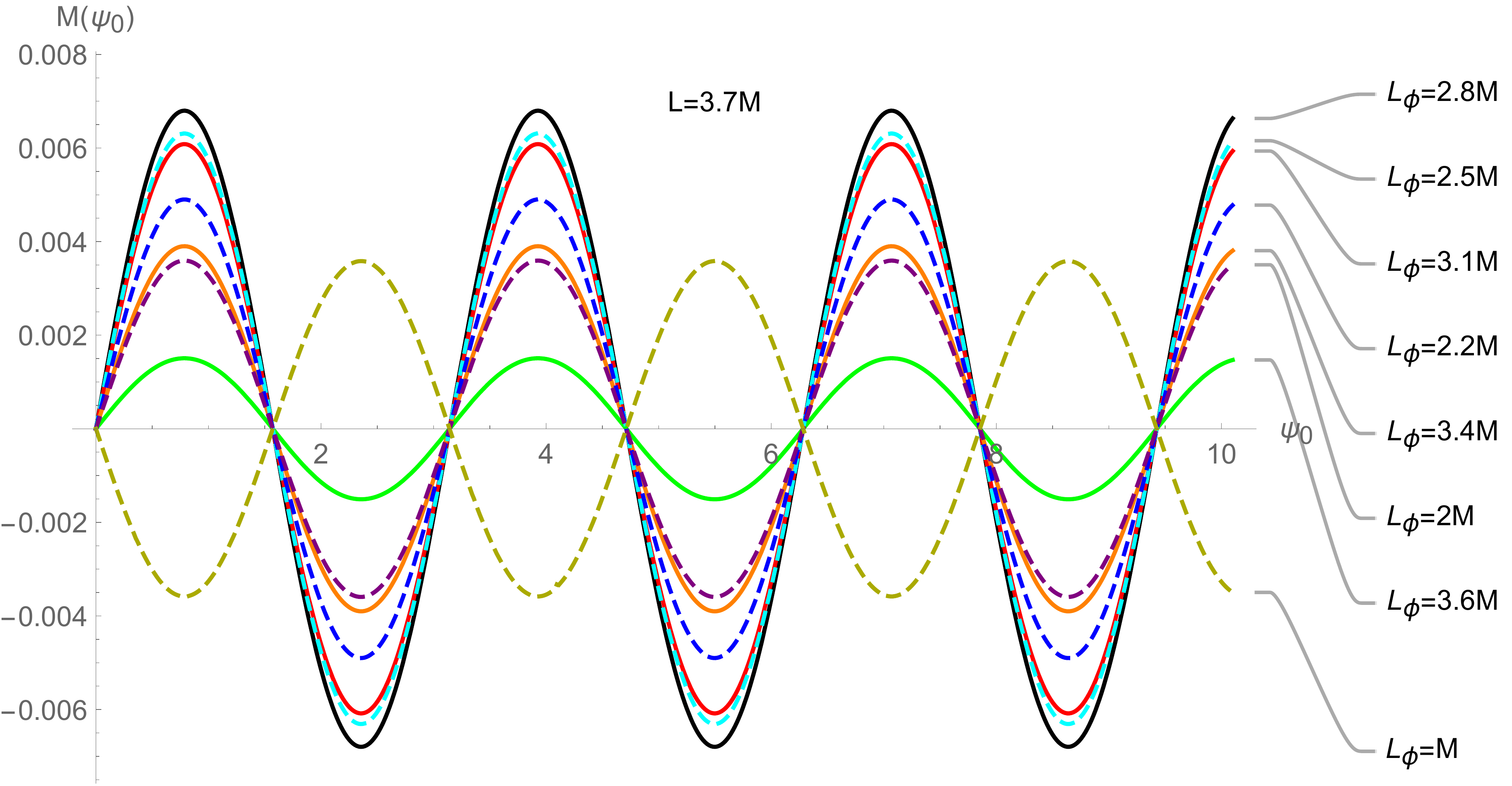}
\includegraphics[width=0.42\textwidth]{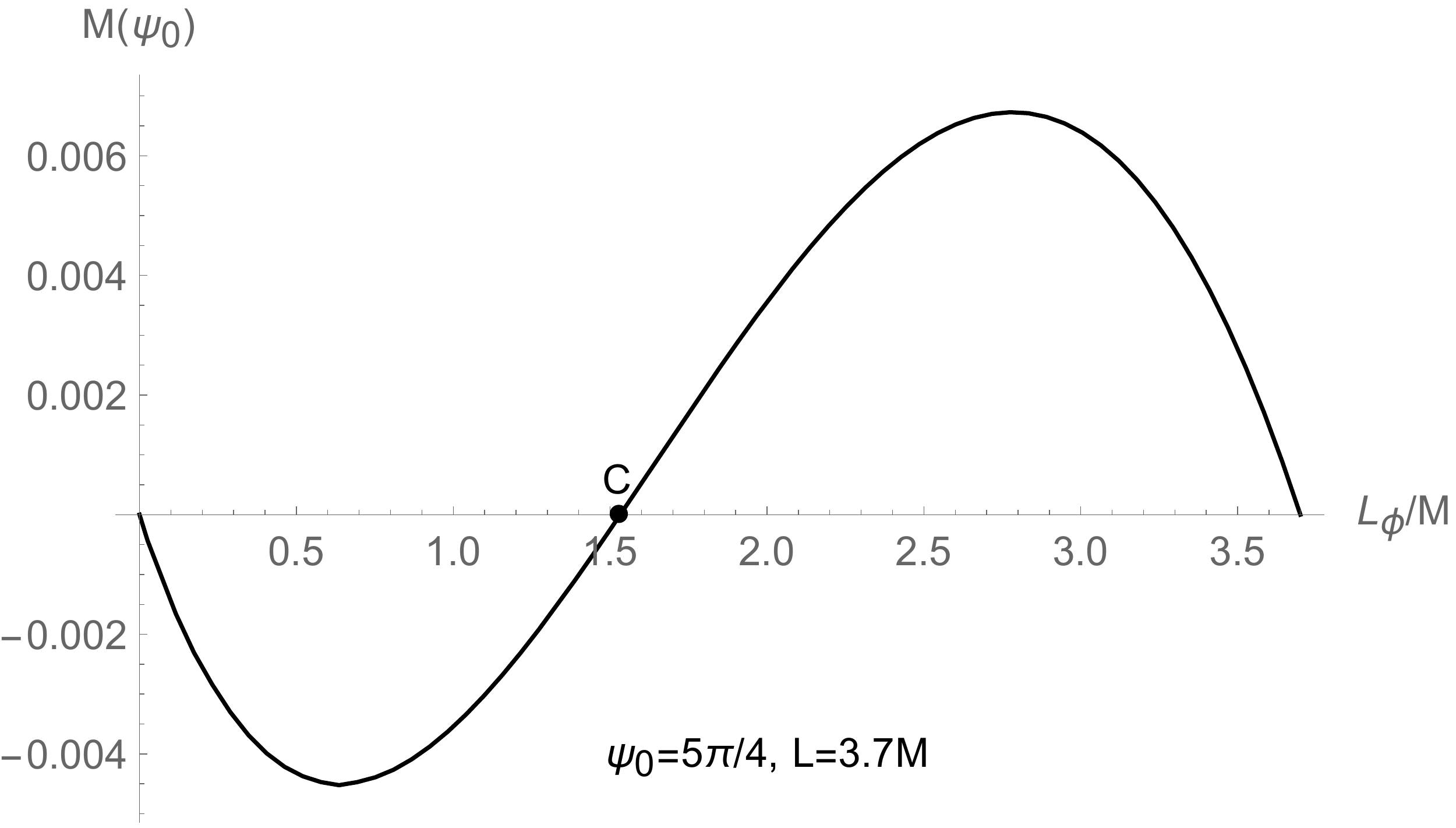}
\end{center}
\caption{{\footnotesize Dependence of the Melnikov function $\mathcal{M}%
\left(  \psi_{0}\right)  $ on the $z$-component of the angular momentum
$L_{\phi}$. We take $m=1$ and $L=3.7M$. \textbf{Left Panel}: $\mathcal{M}%
\left(  \psi_{0}\right)  $ as a function of $\psi_{0}$ for various values of
$L_{\phi}$. $\mathcal{M}\left(  \psi_{0}\right)  $ has simple zeros at the
points $\psi_{0}=n\pi/2$ with $n\in Z$. \textbf{Right Panel}: $\mathcal{M}%
\left(  \psi_{0}=5\pi/4\right)  $ as a function of $L_{\phi}/M$ on the
interval of $0\leq L_{\phi}\leq L$. $\mathcal{M}\left(  \psi_{0}\right)  =0$
when $L_{\phi}=0$, $L_{\phi}=L$ and at the point $C$, where $L_{\phi}%
/M\simeq1.53$. The amplitude of $\mathcal{M}\left(  \psi_{0}\right)  $ has two
local maximum values at $L_{\phi}/M\simeq0.63$ and $2.80$, respectively.}}%
\label{fig:Lphi}%
\end{figure}

The Melnikov function $\mathcal{M}\left(  \psi_{0}\right)  $ is quite complex
and cannot be expressed in closed form. Nevertheless, $\mathcal{M}\left(
\psi_{0}\right)  $ can be computed numerically, and simple zeros of
$\mathcal{M}\left(  \psi_{0}\right)  $ can be observed in its plot. Eqn.
$\left(  \ref{eq:M}\right)  $ shows that, except the prefactor $\beta m^{5}M$,
the behavior of $\mathcal{M}\left(  \psi_{0}\right)  $ only depends on two
dimensionless parameters $L/M$ and $L_{\phi}/M$. So we depict how
$\mathcal{M}\left(  \psi_{0}\right)  $ depends on $L/M$ and $L_{\phi}/M$ in
FIGs. \ref{fig:L} and \ref{fig:Lphi}, respectively. The Melnikov function
$\mathcal{M}\left(  \psi_{0}\right)  $ is plotted for various values of $L/M$
with fixed value of $L_{\phi}/M$ (i.e., $L_{\phi}/M=1$) in the left panel of
FIG. \ref{fig:L}, where we take $m=1$ without loss of generality. As expected,
the periodic function $\mathcal{M}\left(  \psi_{0}\right)  $ has same period
$\pi$ as the periodic coordinate $\psi$. More interestingly, $\mathcal{M}%
\left(  \psi_{0}\right)  $ is shown to have simple zeros at the points
$\psi_{0}=n\pi/2$ with $n\in Z$, which means that the system exhibits chaotic
feature. The panel also shows that the amplitude of $\mathcal{M}\left(
\psi_{0}\right)  $ monotonically grows as the value of $L/M$ increases. This
behavior is also displayed in the right panel, in which the value of
$\mathcal{M}\left(  \psi_{0}\right)  $ at $\psi_{0}=5\pi/4$ is plotted against
$L/M$. In the left panel of FIG. \ref{fig:Lphi}, $\mathcal{M}\left(  \psi
_{0}\right)  $ is plotted for various values of $L_{\phi}/M$ with fixed value
of $L/M$ (i.e., $L/M=3.7$). It also shows that $\mathcal{M}\left(  \psi
_{0}\right)  $ oscillates around zero with $\mathcal{M}\left(  \psi
_{0}\right)  =0$ at $\psi_{0}=n\pi/2$ with $n\in Z$. The Melnikov function
$\mathcal{M}\left(  \psi_{0}\right)  $ is shown to have the maximum amplitude
at $L_{\phi}/M\simeq2.8$. To better illustrate the dependence of the amplitude
of $\mathcal{M}\left(  \psi_{0}\right)  $ on $L_{\phi}/M$, we plot
$\mathcal{M}\left(  \psi_{0}=5\pi/4\right)  $ as a function of $L_{\phi}/M$ in
the right panel of FIG. \ref{fig:Lphi}, in which two local maxima of the
amplitude of $\mathcal{M}\left(  \psi_{0}\right)  $ at $L_{\phi}/M\simeq0.63$
and $2.80$ can be seen. The amplitude of $\mathcal{M}\left(  \psi_{0}\right)
$\ is zero for $L_{\phi}=0$ and $L_{\phi}=L$, which is expected since the
integrand in eqn. $\left(  \ref{eq:M0}\right)  $ is periodic and independent
of $\psi_{0}$ when $L_{\phi}=0$ and $L_{\phi}=L$ (can be seen from eqn.
$\left(  \ref{eq:H0H1}\right)  $). Moreover, it displays that the amplitude of
$\mathcal{M}\left(  \psi_{0}\right)  $\ is also zero for $L_{\phi}%
/M\simeq1.53$. So $\mathcal{M}\left(  \psi_{0}\right)  =0$ for $L_{\phi}=0$,
$L_{\phi}/M=L/M=3.7$ and $L_{\phi}/M\simeq1.53$, which means that the
homoclinic orbit is preserved, and hence there is no occurrence of Smale
horseshoes chaotic motion in these cases.

\section{Conclusion}

\label{Sec:Con}

In this paper, we used the Melnikov method to investigate the chaotic behavior
in geodesic motion on a Schwarzschild metric perturbed by the minimal length
effects. The unperturbed system is well known to be integrable. For the near
integrable perturbed system, the Melnikov method is very powerful to detect
the presence of chaotic structure by tracing simple zeros of the Melnikov
function $\mathcal{M}\left(  \psi_{0}\right)  $. After the perturbed
Hamiltonian for a massive particle of angular momentum per unit mass $L$ and
$z$-component of the orbital angular momentum per unit mass $L_{\phi}$ was
obtained, the Melnikov function $\mathcal{M}\left(  \psi_{0}\right)  $ was
numerically evaluated by using the higher-dimensional generalization of the
Melnikov method. We make three observations regarding $\mathcal{M}\left(
\psi_{0}\right)  $:

\begin{itemize}
\item When $L_{\phi}=0$, $L$ and $L_{C}$ with $0<L_{C}<L$, $\mathcal{M}\left(
\psi_{0}\right)  =0$, which implies that no Smale horseshoes chaotic motion is
present in the perturbed system.

\item When the amplitude of $\mathcal{M}\left(  \psi_{0}\right)  $ is not
zero, $\mathcal{M}\left(  \psi_{0}\right)  $ is a periodic function with the
period of $\pi$, and has simple zeros at $\psi_{0}=n\pi/2$ with $n\in Z$,
which signals the appearance of Smale horseshoes chaotic structure in the
perturbed system.

\item The amplitude of $\mathcal{M}\left(  \psi_{0}\right)  $ increases as $L$
increases with fixed $L_{\phi}$. When $L$ is fixed, the amplitude of
$\mathcal{M}\left(  \psi_{0}\right)  $ as a function of $L_{\phi}$ has two
local maxima.
\end{itemize}

The Melnikov's method provides necessary but not sufficient condition for
chaos and serves as an independent check on numerical tests for chaos. So it
would be interesting to use other chaos indicators, e.g., the Poincare
surfaces of section, the Lyapunov characteristic exponents and the method of
fractal basin boundaries, to detect chaotic behavior in systems perturbed by
the minimal length effects. In \cite{IN-Lu:2018mpr}, we calculated the minimal
length effects on the Lyapunov exponent of a massive particle perturbed away
from an unstable equilibrium near the black hole horizon and found that the
classical trajectory in black holes becomes more chaotic, which is consistent
with the chaotic behavior found in this paper. Finally, the minimal length
effects on the dual conformal field theory was analyzed in
\cite{CON-Faizal:2017dlb}. It is tempting to understand the holographic
aspects of these chaotic behavior.

\begin{acknowledgments}
We are grateful to Houwen Wu and Haitang Yang for useful discussions. This
work is supported in part by NSFC (Grant No. 11875196, 11375121 and 11005016),
the Fundamental Research Funds for the Central Universities, Natural Science
Foundation of Chengdu University of TCM (Grants nos. ZRYY1729 and ZRYY1921),
Discipline Talent Promotion Program of /Xinglin Scholars(Grant no.
QNXZ2018050) and the key fund project for Education Department of Sichuan
(Grant no. 18ZA0173).
\end{acknowledgments}


\noindent

\begin{thebibliography}{99}                                                                                               %


\bibitem {IN-Carter:1968rr}B.~Carter, \textquotedblleft Global structure of
the Kerr family of gravitational fields,\textquotedblright%
\ Phys.\ Rev.\ \textbf{174}, 1559 (1968). doi:10.1103/PhysRev.174.1559

\bibitem {IN-Sota:1995ms}Y.~Sota, S.~Suzuki and K.~i.~Maeda, ``Chaos in static
axisymmetric space-times. 1: Vacuum case,'' Class.\ Quant.\ Grav.\ \textbf{13}%
, 1241 (1996) doi:10.1088/0264-9381/13/5/034 [gr-qc/9505036].

\bibitem {IN-Hanan:2006uf}W.~Hanan and E.~Radu, ``Chaotic motion in
multi-black hole spacetimes and holographic screens,'' Mod.\ Phys.\ Lett.\ A
\textbf{22}, 399 (2007) doi:10.1142/S0217732307022815 [gr-qc/0610119].

\bibitem {IN-Gair:2007kr}J.~R.~Gair, C.~Li and I.~Mandel, ``Observable
Properties of Orbits in Exact Bumpy Spacetimes,'' Phys.\ Rev.\ D \textbf{77},
024035 (2008) doi:10.1103/PhysRevD.77.024035 [arXiv:0708.0628 [gr-qc]].

\bibitem {IN-Witzany:2015yqa}V.~Witzany, O.~Semer\'{a}k and P.~Sukov\'{a},
``Free motion around black holes with discs or rings: between integrability
and chaos -- IV,'' Mon.\ Not.\ Roy.\ Astron.\ Soc.\ \textbf{451}, no. 2, 1770
(2015) doi:10.1093/mnras/stv1148 [arXiv:1503.09077 [astro-ph.HE]].

\bibitem {IN-Wang:2016wcj}M.~Wang, S.~Chen and J.~Jing, ``Chaos in the motion
of a test scalar particle coupling to the Einstein tensor in
Schwarzschild--Melvin black hole spacetime,'' Eur.\ Phys.\ J.\ C \textbf{77},
no. 4, 208 (2017) doi:10.1140/epjc/s10052-017-4792-y [arXiv:1605.09506 [gr-qc]].

\bibitem {IN-Chen:2016tmr}S.~Chen, M.~Wang and J.~Jing, ``Chaotic motion of
particles in the accelerating and rotating black holes spacetime,'' JHEP
\textbf{1609}, 082 (2016) doi:10.1007/JHEP09(2016)082 [arXiv:1604.02785 [gr-qc]].

\bibitem {IN-Liu:2018bmn}C.~Y.~Liu, \textquotedblleft Chaotic Motion of
Charged Particles around a Weakly Magnetized Kerr-Newman Black
Hole,\textquotedblright\ arXiv:1806.09993 [gr-qc].

\bibitem {IN-Frolov:1999pj}A.~V.~Frolov and A.~L.~Larsen, ``Chaotic scattering
and capture of strings by black hole,'' Class.\ Quant.\ Grav.\ \textbf{16},
3717 (1999) doi:10.1088/0264-9381/16/11/316 [gr-qc/9908039].

\bibitem {IN-Zayas:2010fs}L.~A.~Pando Zayas and C.~A.~Terrero-Escalante,
``Chaos in the Gauge / Gravity Correspondence,'' JHEP \textbf{1009}, 094
(2010) doi:10.1007/JHEP09(2010)094 [arXiv:1007.0277 [hep-th]].

\bibitem {IN-Ma:2014aha}D.~Z.~Ma, J.~P.~Wu and J.~Zhang, \textquotedblleft
Chaos from the ring string in a Gauss-Bonnet black hole in AdS5
space,\textquotedblright\ Phys.\ Rev.\ D \textbf{89}, no. 8, 086011 (2014)
doi:10.1103/PhysRevD.89.086011 [arXiv:1405.3563 [hep-th]].

\bibitem {IN-Ma:2019ewq}D.~Z.~Ma, D.~Zhang, G.~Fu and J.~P.~Wu,
\textquotedblleft Chaotic dynamics of string around charged black brane with
hyperscaling violation,\textquotedblright\ JHEP \textbf{2001}, 103 (2020)
doi:10.1007/JHEP01(2020)103 [arXiv:1911.09913 [hep-th]].

\bibitem {IN-Mel}V. K. Mel'nikov, \textquotedblleft On the stability of a
center for time-periodic perturbations\textquotedblright, Tr. Mosk. Mat. Obs.,
12, 1963, 3--52

\bibitem {IN-Bombelli:1991eg}L.~Bombelli and E.~Calzetta, ``Chaos around a
black hole,'' Class.\ Quant.\ Grav.\ \textbf{9}, 2573 (1992). doi:10.1088/0264-9381/9/12/004

\bibitem {IN-Letelier:1996he}P.~S.~Letelier and W.~M.~Vieira, ``Chaos in black
holes surrounded by gravitational waves,'' Class.\ Quant.\ Grav.\ \textbf{14},
1249 (1997) doi:10.1088/0264-9381/14/5/026 [gr-qc/9706025].

\bibitem {IN-Santoprete:2001wz}M.~Santoprete and G.~Cicogna, ``Chaos in black
holes surrounded by electromagnetic fields,'' Gen.\ Rel.\ Grav.\ \textbf{34},
1107 (2002) doi:10.1023/A:1016570106387 [nlin/0110046 [nlin-cd]].

\bibitem {IN-Polcar:2019kwu}L.~Polcar and O.~Semer\'{a}k, ``Free motion around
black holes with discs or rings: Between integrability and chaos. VI. The
Melnikov method,'' Phys.\ Rev.\ D \textbf{100}, no. 10, 103013 (2019)
doi:10.1103/PhysRevD.100.103013 [arXiv:1911.09790 [gr-qc]].

\bibitem {IN-Chabab:2018lzf}M.~Chabab, H.~El Moumni, S.~Iraoui, K.~Masmar and
S.~Zhizeh, ``Chaos in charged AdS black hole extended phase space,''
Phys.\ Lett.\ B \textbf{781}, 316 (2018) doi:10.1016/j.physletb.2018.04.014
[arXiv:1804.03960 [hep-th]].

\bibitem {IN-Mahish:2019tgv}S.~Mahish and B.~Chandrasekhar, ``Chaos in Charged
Gauss-Bonnet AdS Black Holes in Extended Phase Space,'' Phys.\ Rev.\ D
\textbf{99}, no. 10, 106012 (2019) doi:10.1103/PhysRevD.99.106012
[arXiv:1902.08932 [hep-th]].

\bibitem {IN-Chen:2019bwt}Y.~Chen, H.~Li and S.~J.~Zhang, ``Chaos in
Born--Infeld--AdS black hole within extended phase space,''
Gen.\ Rel.\ Grav.\ \textbf{51}, no. 10, 134 (2019)
doi:10.1007/s10714-019-2612-4 [arXiv:1907.08734 [hep-th]].

\bibitem {IN-Dai:2020wny}C.~Dai, S.~Chen and J.~Jing, \textquotedblleft
Thermal chaos of a charged dilaton-AdS black hole in the extended phase
space,\textquotedblright\ arXiv:2002.01641 [gr-qc].

\bibitem {IN-Veneziano:1986zf}G.~Veneziano, \textquotedblleft A Stringy Nature
Needs Just Two Constants,\textquotedblright\ Europhys.\ Lett.\ \textbf{2}, 199
(1986). doi:10.1209/0295-5075/2/3/006

\bibitem {IN-Gross:1987ar}D.~J.~Gross and P.~F.~Mende, ``String Theory Beyond
the Planck Scale,'' Nucl.\ Phys.\ B \textbf{303}, 407 (1988). doi:10.1016/0550-3213(88)90390-2

\bibitem {IN-Amati:1988tn}D.~Amati, M.~Ciafaloni and G.~Veneziano, ``Can
Space-Time Be Probed Below the String Size?,'' Phys.\ Lett.\ B \textbf{216},
41 (1989). doi:10.1016/0370-2693(89)91366-X

\bibitem {IN-Garay:1994en}L.~J.~Garay, ``Quantum gravity and minimum length,''
Int.\ J.\ Mod.\ Phys.\ A \textbf{10}, 145 (1995) doi:10.1142/S0217751X95000085 [gr-qc/9403008].

\bibitem {IN-Maggiore:1993kv}M.~Maggiore, ``The Algebraic structure of the
generalized uncertainty principle,'' Phys.\ Lett.\ B \textbf{319}, 83 (1993)
doi:10.1016/0370-2693(93)90785-G [hep-th/9309034].

\bibitem {IN-Kempf:1994su}A.~Kempf, G.~Mangano and R.~B.~Mann, ``Hilbert space
representation of the minimal length uncertainty relation,'' Phys.\ Rev.\ D
\textbf{52}, 1108 (1995) doi:10.1103/PhysRevD.52.1108 [hep-th/9412167].

\bibitem {IN-Chang:2001kn}L.~N.~Chang, D.~Minic, N.~Okamura and T.~Takeuchi,
``Exact solution of the harmonic oscillator in arbitrary dimensions with
minimal length uncertainty relations,'' Phys.\ Rev.\ D \textbf{65}, 125027
(2002) doi:10.1103/PhysRevD.65.125027 [hep-th/0111181].

\bibitem {IN-Akhoury:2003kc}R.~Akhoury and Y.~P.~Yao, ``Minimal length
uncertainty relation and the hydrogen spectrum,'' Phys.\ Lett.\ B
\textbf{572}, 37 (2003) doi:10.1016/j.physletb.2003.07.084 [hep-ph/0302108].

\bibitem {IN-Brau:1999uv}F.~Brau, \textquotedblleft Minimal length uncertainty
relation and hydrogen atom,\textquotedblright\ J.\ Phys.\ A \textbf{32}, 7691
(1999) doi10.1088/0305-4470/32/44/308 [quant-ph/9905033].

\bibitem {IN-Brau:2006ca}F.~Brau and F.~Buisseret, ``Minimal Length
Uncertainty Relation and gravitational quantum well,'' Phys.\ Rev.\ D
\textbf{74}, 036002 (2006) doi:10.1103/PhysRevD.74.036002 [hep-th/0605183].

\bibitem {IN-Pedram:2011xj}P.~Pedram, K.~Nozari and S.~H.~Taheri,
\textquotedblleft The effects of minimal length and maximal momentum on the
transition rate of ultra cold neutrons in gravitational
field,\textquotedblright\ JHEP \textbf{1103}, 093 (2011)
doi:10.1007/JHEP03(2011)093 [arXiv:1103.1015 [hep-th]].

\bibitem {IN-Pikovski:2011zk}I.~Pikovski, M.~R.~Vanner, M.~Aspelmeyer,
M.~S.~Kim and C.~Brukner, ``Probing Planck-scale physics with quantum
optics,'' Nature Phys.\ \textbf{8}, 393 (2012) doi:10.1038/nphys2262
[arXiv:1111.1979 [quant-ph]].

\bibitem {IN-Bosso:2018ckz}P.~Bosso, S.~Das and R.~B.~Mann, ``Potential tests
of the Generalized Uncertainty Principle in the advanced LIGO experiment,''
Phys.\ Lett.\ B \textbf{785}, 498 (2018) doi:10.1016/j.physletb.2018.08.061
[arXiv:1804.03620 [gr-qc]].

\bibitem {IN-Wang:2010ct}P.~Wang, H.~Yang and X.~Zhang, ``Quantum gravity
effects on statistics and compact star configurations,'' JHEP \textbf{1008},
043 (2010) doi:10.1007/JHEP08(2010)043 [arXiv:1006.5362 [hep-th]].

\bibitem {IN-Ong:2018zqn}Y.~C.~Ong, \textquotedblleft Generalized Uncertainty
Principle, Black Holes, and White Dwarfs: A Tale of Two
Infinities,\textquotedblright\ JCAP \textbf{1809}, no. 09, 015 (2018)
doi:10.1088/1475-7516/2018/09/015 [arXiv:1804.05176 [gr-qc]].

\bibitem {IN-Benczik:2002tt}S.~Benczik, L.~N.~Chang, D.~Minic, N.~Okamura,
S.~Rayyan and T.~Takeuchi, \textquotedblleft Short distance versus long
distance physicsThe Classical limit of the minimal length uncertainty
relation,\textquotedblright\ Phys.\ Rev.\ D \textbf{66}, 026003 (2002)
doi10.1103/PhysRevD.66.026003 [hep-th/0204049].

\bibitem {IN-Silagadze:2009vu}Z.~K.~Silagadze, \textquotedblleft Quantum
gravity, minimum length and Keplerian orbits,\textquotedblright%
\ Phys.\ Lett.\ A \textbf{373}, 2643 (2009) doi:10.1016/j.physleta.2009.05.053
[arXiv:0901.1258 [gr-qc]].

\bibitem {IN-Ahmadi:2014cga}F.~Ahmadi and J.~Khodagholizadeh,
\textquotedblleft Effect of GUP on the Kepler problem and a variable minimal
length,\textquotedblright\ Can.\ J.\ Phys.\ \textbf{92}, 484 (2014)
doi:10.1139/cjp-2013-0354 [arXiv:1411.0241 [hep-th]].

\bibitem {IN-Scardigli:2014qka}F.~Scardigli and R.~Casadio, ``Gravitational
tests of the Generalized Uncertainty Principle,'' Eur.\ Phys.\ J.\ C
\textbf{75}, no. 9, 425 (2015) doi:10.1140/epjc/s10052-015-3635-y
[arXiv:1407.0113 [hep-th]].

\bibitem {IN-Ali:2015zua}A.~Farag Ali, M.~M.~Khalil and E.~C.~Vagenas,
\textquotedblleft Minimal Length in quantum gravity and gravitational
measurements,\textquotedblright\ Europhys.\ Lett.\ \textbf{112}, no. 2, 20005
(2015) doi:10.1209/0295-5075/112/20005 [arXiv:1510.06365 [gr-qc]].

\bibitem {IN-Guo:2015ldd}X.~Guo, P.~Wang and H.~Yang, \textquotedblleft The
classical limit of minimal length uncertainty relation: revisit with the
Hamilton-Jacobi method,\textquotedblright\ JCAP \textbf{1605}, no. 05, 062
(2016) doi:10.1088/1475-7516/2016/05/062 [arXiv:1512.03560 [gr-qc]].

\bibitem {IN-Khodadi:2017eim}M.~Khodadi, K.~Nozari and A.~Hajizadeh,
\textquotedblleft Some Astrophysical Aspects of a Schwarzschild Geometry
Equipped with a Minimal Measurable Length,\textquotedblright\ Phys.\ Lett.\ B
\textbf{770}, 556 (2017) doi:10.1016/j.physletb.2017.05.016 [arXiv:1702.06357 [gr-qc]].

\bibitem {IN-Scardigli:2018qce}F.~Scardigli and R.~Casadio, ``Perihelion
Precession and Generalized Uncertainty Principle,'' Springer
Proc.\ Phys.\ \textbf{208}, 149 (2018).

\bibitem {IN-Tao:2012fp}J.~Tao, P.~Wang and H.~Yang, \textquotedblleft
Homogeneous Field and WKB Approximation In Deformed Quantum Mechanics with
Minimal Length,\textquotedblright\ Adv.\ High Energy Phys.\ \textbf{2015},
718359 (2015) doi:10.1155/2015/718359 [arXiv:1211.5650 [hep-th]].

\bibitem {IN-Quintela:2015bua}T.~S.~Quintela, Jr., J.~C.~Fabris and
J.~A.~Nogueira, ``The Harmonic Oscillator in the Classical Limit of a
Minimal-Length Scenario,'' Braz.\ J.\ Phys.\ \textbf{46}, no. 6, 777 (2016)
doi:10.1007/s13538-016-0457-9 [arXiv:1510.08129 [hep-th]].

\bibitem {IN-Tkachuk:2013qa}V.~M.~Tkachuk, \textquotedblleft Deformed
Heisenberg algebra with minimal length and equivalence
principle,\textquotedblright\ Phys.\ Rev.\ A \textbf{86}, 062112 (2012)
doi:10.1103/PhysRevA.86.062112 [arXiv:1301.1891 [gr-qc]].

\bibitem {IN-Scardigli:2016pjs}F.~Scardigli, G.~Lambiase and E.~Vagenas, ``GUP
parameter from quantum corrections to the Newtonian potential,''
Phys.\ Lett.\ B \textbf{767}, 242 (2017) doi:10.1016/j.physletb.2017.01.054
[arXiv:1611.01469 [hep-th]].

\bibitem {IN-Zhao:2017xjj}Q.~Zhao, M.~Faizal and Z.~Zaz, ``Short distance
modification of the quantum virial theorem,'' Phys.\ Lett.\ B \textbf{770},
564 (2017) doi:10.1016/j.physletb.2017.01.029 [arXiv:1707.00636 [hep-th]].

\bibitem {IN-Mu:2019bim}B.~Mu and J.~Tao, \textquotedblleft Minimal Length
Effect on Thermodynamics and Weak Cosmic Censorship Conjecture in anti-de
Sitter Black Holes via Charged Particle Absorption,\textquotedblright%
\ arXiv:1906.10544 [gr-qc].

\bibitem {IN-Lu:2018mpr}F.~Lu, J.~Tao and P.~Wang, \textquotedblleft Minimal
Length Effects on Chaotic Motion of Particles around Black Hole
Horizon,\textquotedblright\ JCAP \textbf{1812}, 036 (2018)
doi:10.1088/1475-7516/2018/12/036 [arXiv:1811.02140 [gr-qc]].

\bibitem {IN-Hassanabadi:2019iff}H.~Hassanabadi, E.~Maghsoodi and W.~Sang
Chung, ``Analysis of motion of particles near black hole horizon under
generalized uncertainty principle,'' EPL \textbf{127}, no. 4, 40002 (2019). doi:10.1209/0295-5075/127/40002

\bibitem {IN-Chen:2013tha}D.~Chen, H.~Wu and H.~Yang, ``Observing remnants by
fermions' tunneling,'' JCAP \textbf{1403}, 036 (2014)
doi:10.1088/1475-7516/2014/03/036 [arXiv:1307.0172 [gr-qc]].

\bibitem {IN-Chen:2013ssa}D.~Y.~Chen, Q.~Q.~Jiang, P.~Wang and H.~Yang,
\textquotedblleft Remnants, fermions` tunnelling and effects of quantum
gravity,\textquotedblright\ JHEP \textbf{1311}, 176 (2013)
doi:10.1007/JHEP11(2013)176 [arXiv:1312.3781 [hep-th]].

\bibitem {IN-Chen:2014xgj}D.~Chen, H.~Wu, H.~Yang and S.~Yang,
\textquotedblleft Effects of quantum gravity on black holes,\textquotedblright%
\ Int.\ J.\ Mod.\ Phys.\ A \textbf{29}, no. 26, 1430054 (2014)
doi:10.1142/S0217751X14300543 [arXiv:1410.5071 [gr-qc]].

\bibitem {IN-Maghsoodi:2019fca}E.~Maghsoodi, H.~Hassanabadi and W.~Sang Chung,
``Black hole thermodynamics under the generalized uncertainty principle and
doubly special relativity,'' PTEP \textbf{2019}, no. 8, 083E03 (2019)
doi:10.1093/ptep/ptz085 [arXiv:1901.10305 [physics.gen-ph]].

\bibitem {MM-Wiggins}S. Wiggins, \textquotedblleft Introduction to Applied
Nonlinear Dynamical Systems and Chaos (Second Ed.),\textquotedblright%
\ Springer-Verlag, New York and Bristol, 2003.

\bibitem {MM-HM}P. J. Holmes and J. E. Marsden, \textquotedblleft Horseshoes
and Arnold diffusion for Hamiltonian Systems on Lie groups,\textquotedblright%
\ Indiana Univ. Math. J. 32, 273 (1983).

\bibitem {CUMLE-Mu:2015qta}B.~Mu, P.~Wang and H.~Yang, ``Minimal Length
Effects on Tunnelling from Spherically Symmetric Black Holes,'' Adv.\ High
Energy Phys.\ \textbf{2015}, 898916 (2015) doi:10.1155/2015/898916
[arXiv:1501.06025 [gr-qc]].

\bibitem {CON-Faizal:2017dlb}M.~Faizal, A.~F.~Ali and A.~Nassar, ``Generalized
uncertainty principle as a consequence of the effective field theory,''
Phys.\ Lett.\ B \textbf{765}, 238 (2017) doi:10.1016/j.physletb.2016.11.054
[arXiv:1701.00341 [hep-th]].
\end{thebibliography}
\end{document}